\documentclass[a4paper,twocolumn]{article}

\usepackage{times}
\usepackage{graphicx}
\usepackage[authoryear]{natbib}

\usepackage{amssymb}
\usepackage{epstopdf}

\begin{document}

\title{Galaxy morphology - an unsupervised machine learning approach}

\author{\parbox{\textwidth}{\flushleft
\vspace{-0.5cm}
{Andrew Schutter, Lior Shamir$^*$}\\
\vspace{0.4cm}
{\small Lawrence Technological University, Southfield, Michigan 48075, USA} 
}}

\date{}

\maketitle

\begin{abstract}
Structural properties posses valuable information about the formation and evolution of galaxies, and are important for understanding the past, present, and future universe. Here we use unsupervised machine learning methodology to analyze a network of similarities between galaxy morphological types, and automatically deduce a morphological sequence of galaxies. Application of the method to the EFIGI catalog show that the morphological scheme produced by the algorithm is largely in agreement with the De Vaucouleurs system, demonstrating the ability of computer vision and machine learning methods to automatically profile galaxy morphological sequences. The unsupervised analysis method is based on comprehensive computer vision techniques that compute the visual similarities between the different morphological types. Rather than relying on human cognition, the proposed system deduces the similarities between sets of galaxy images in an automatic manner, and is therefore not limited by the number of galaxies being analyzed. The source code of the method is publicly available, and the protocol of the experiment is included in the paper so that the experiment can be replicated, and the method can be used to analyze user-defined datasets of galaxy images.
\end{abstract}

{\bf Keywords:}  
galaxies: structure -- galaxies: evolution -- methods: analytical -- techniques: image processing \newline \newline

\section{Introduction}
\label{introduction}

In the past few years, advancements in computational tools and algorithms have started to allow automatic analysis of galaxy morphology. Approaches to automatic galaxy classification include model-driven methods such as GALFIT \citep{pen02}, GIM2D \citep{sim99, sim11}, CAS \citep{con03}, Gini\citep{Abr03}, Ganalyzer \citep{sha11}, and SpArcFiRe \citep{davis2014sparcfire}. Data-driven methods include binary classifiers that can differentiate between broad galaxy morphological types of elliptical and spiral galaxies \citep{sha09,meneses2009classification,banerji2010}, but also classifiers that can differentiate between four basic objects \citep{abd2013intelligent}, classification between four basic Hubble morphological types of E, S0, Sab, and Scd \citep{hue10}, and comprehensive analysis of galaxy images that include specific morphological features \citep{baillard2006project,kum14,dieleman2015rotation}. Classification of galaxies can also be performed using spectra in supervised \citep{ball2004} and  unsupervised \citep{almeida2010automatic} manner.

While supervised machine learning have demonstrated reasonable efficacy in automatic classification of galaxies by their morphological types \citep{sha09,meneses2009classification,banerji2010,hue10}, discrete classifiers do not effectively conceptualize the continuous nature of galaxy morphology, and therefore galaxy morphological schemes are still defined by manual observation. One of the earlier and most widely used schemes is the Hubble sequence \citep{hubble36,san61}, which is a commonly used morphology classification scheme that covers the morphology of most known galaxies. Hubble's initial work proposed a morphology classification system based on attributes of observed nebulae, originally consisting of three main morphological types, commonly known as elliptical (E), normal spirals (S) and barred spirals (SB) \citep{hubble36}. \cite{humason1956redshifts} revisited the Hubble Sequence, introducing lenticular galaxies (S0), creating what is most commonly known as the Hubble ``tuning-fork'' diagram \citep{de59}.  It should be noted that although irregular (I) galaxies were recognized by Hubble, they were not included in Hubble's classification scheme since at the time they could not be distinctively classified \citep{de59}.

Since the Hubble morphological scheme was introduced, several modifications and enhancements have been proposed. \cite{Mor57},  proposed a galaxy classification scheme based on the spectra, showing the correlation between the spectra and the spiral structure and spectral concentration \citep{Mor58}, and identified the cD phenomena \citep{mor65}. \cite{van60} proposed a classification system of late-type galaxies based on luminosity, driven by the correlation between absolute luminosity and the shape of the spiral arms. That work was followed by a galaxy classification scheme of spiral and S0 galaxies, and distinguished ``early'' and ``late'' type systems by their disk-to-bulge ratio \citep{van76}.

\cite{san61} showed that S0$_{1}$ to S0$_{3}$ galaxies do not feature an increase in flattening of the galaxies, and that normal spiral galaxies and S0 galaxies form two parallel sequences \citep{san70,van76}.  \cite{kor96} expanded the Hubble classification scheme with a more detailed morphological analysis of elliptical galaxies \citep{kor89}. \cite{kor96} proposed some modifications to the Hubble sequence, including the two-component S0 galaxies, and the addition of the Magellanic irregulars.

One of the notable refinements and extensions to the Hubble sequence was proposed by \cite{de59}, proposing a three dimensional system.  This classification included the four main broad morphological classes of elliptical, lenticular, spiral, and irregular galaxies along a linear main axis from galaxy types E to I$_{m}$, including Hubble's initial a, b, c representation for "early" to "late". \cite{san61} refinement included d for "very late" and the division of S0 galaxies into SO$^{-}$, SO$^{0}$, SO$^{+}$, as well as the inclusion of "m" for magellanic galaxies: E, E$^{+}$, S0$^{-}$, S0$^{0}$, S0$^{+}$, S$_{a}$, S$_{b}$, S$_{c}$, S$_{d}$, S$_{m}$, or I$_{m}$ \citep{de1994global}. The classification was also extended to include intermediate stages between the initial a, b, c, d, and m stages such as ab, bc, cd, and dm.  This scheme introduced a notation based on \textit{family}, \textit{variety}, and \textit{stage}, with \textit{family} representing the absence of bars in a spiral galaxy (A), the presence of bars (B), or a transition of the two (AB). \textit{Variety} represents the presence of a ring shape (r), spiral shape (s), or transition of the two (rs) within spiral galaxies, and \textit{stage} represents the galaxy position along the main axis. Another feature of this classification scheme was assigning each \textit{stage} along the main axis a numerical integer value between -6 and 11.  E galaxies being represented by the values -6 to -4, lenticular -3 to -1, sprials 0 to 9, and irregulars 10 to 11 for a more quantitative approach to the classification.  Furthering the quantitative approach to galaxy classification, \cite{de1994global} also introduced measurable parameters showing either a consistent mean increase or decrease along the current classification sequence. Characteristics include bulge-to-disk ratios, integrated luminosity in the B-band, the ratio of aperture diameters, total or effective magnitudes, mean surface brightness, and hydrogen index \citep{de1994global}.

While proposing a quantitative scheme, the association of a galaxy to a morphological type is subjective, and the annotations of two or more astronomers are not necessarily identical in all cases \citep{naim1995comparative, de11}. It has been therefore proposed that galaxy morphology classification schemes will involve computational methods \citep{de1994global}. Here we perform automatic unsupervised analysis of galaxy images of different morphological types to produce a computer-generated galaxy morphology sequence. The scheme is based on quantitative computer analysis of thousands of annotated galaxy images, producing a network of similarities between the morphological types that is independent of the human perception and the way humans quantify the similarities between these types.

\section{Data}
\label{data}

The data used in the study are taken from the EFIGI catalog \citep{efigi,de11}, which was compiled for the purpose of developing and testing computational methods related to galaxy morphology. The catalog contains image data as well as morphological annotation data of 4458 galaxies taken from PGC (Principal Galaiesy Catalogue), also included in SDSS (Sloan Digital Sky Survey) Data Release 4. Among other morphological features, each galaxy was assigned with its  morphological type determined by 10 astronomers \citep{efigi} based on the updated RC3-based Hubble types \citep{rc3}. Other morphological features include the bulge, spiral arms, as well as other features such as texture, appearance in the sky, and environment.

EFIGI contains images of each galaxy in the u, g, i, r, and z bands \citep{efigi}. To produce color images,  the i, r, and g bands were combined to provide a composite RGB image, such that gamma correction of 1.3 was applied to the luminosity, and color saturation was increased by a factor of 2. The color images were converted to the  PNG (Portable Network Graphics) format using the STIFF software \citep{efigi}.

The EFIGI color images were converted to 255$\times$255 color 24-bit TIFF (Tagged Image File Format) images using {\it ImageMagick}, and were separated into folders such that each folders contained galaxies of the same type as annotated by EFIGI. Images of the same galaxies in the u, g, r, i, and z bands were converted to monochrome TIFF, and were used without color information.

The galaxy types are based on the numerical scheme \citep{de59} taken from the EFIGI catalog \citep{efigi}. Each galaxy type had at least 142 samples, except for cE (-6), cD (-4), and dE (11), which only had 18, 44, and 69 samples, respectively. For their small size, these classes were not used in the experiment.

\section{Image analysis method}
\label{method}

The image analysis method used in the experiment is Wndchrm \citep{Sha08,shamir2008evaluation,shamir2009knee,shamir2010impressionism,shamir2013automatic}, that has a feature set of 4027 numerical image content descriptors, or 2885 numerical descriptors when color information is not used. The numerical image content descriptors are the following: \newline \newline
Texture features: \newline
1. {\bf Haralick textures}: Energy and entropy computed on the co-occurrence matrix of the image \citep{haralick1973textural}, measured using 28 image descriptor values as described in \cite{Sha08}. \newline
2. {\bf Tamura textures}: {\it Contrast}, {\it directionality} and {\it coarseness} of the Tamura textures \citep{tamura1978textural}. The coarseness descriptors are its sum and its 3-bin histogram, providing a total of six numerical content descriptors. \newline
3. {\bf Gabor Filters}: Gabor filters \citep{gabor1946theory} using seven frequencies (1 through 7) and Gaussian harmonic function \citep{grigorescu2002comparison}. \newline \newline
Polynomial decomposition: \newline
1. {\bf Radon transform features} \citep{Lim90}: Four series computed for angles 0, 45, 90, 135 degrees, and then convolved into a 3-bin histogram, providing a total of 12 numerical content descriptors. \newline
2. {\bf Chebyshev Statistics} \citep{Gra94}: A 32-bin histogram of a 400-bin vector produced by the Chebyshev transform of the with order of N=20. \newline
3. {\bf Zernike features}: Absolute values of the 72 coefficients of the Zernike polynomial approximation \citep{teague1980image}. \newline
4. {\bf Chebyshev-Fourier features}: A 32-bin histogram of the polynomial coefficients of a Chebyshev--Fourier transform \citep{orlov2008wnd} with maximum polynomial order of N=23. \newline \newline
High-contrast features: \newline
1. {\bf Fractal features}, as described in \citep{Wu92}. \newline
2. {\bf Edge features}:  Mean, median, variance, and 8-bin histogram of the magnitude and direction computed on the Prewitt gradient of the image, as well as edge direction homogeneity. \newline
3. {\bf High-contrast object statistics}: Minimum, maximum, mean, median, variance, Euler number, and 10-bin histogram of the objects areas computed on the 8-connected objects found in the Otsu binary transform of the image. \newline \newline
Pixel statistics: \newline
1. {\bf Multi-scale Histograms}: Four histograms with 3, 5, 7, and 9 bins computed on the pixel intensities \citep{Had01}. \newline
2. {\bf First 4 Moments}: Mean, standard deviation, skewness, and kurtosis computed on image "stripes'' in four different directions (0, 45, 90, and 135 degrees). \newline

These features are extracted not just from the raw values, but also from the two-dimensional transforms and combinations of multi-order transforms. The transforms are Fourier transform, Chebyshev transform, Wavelet (symlet 5, level 1) transform, color transform \citep{shamir2006human}, and edge magnitude transform. A detailed description and performance analysis of the image features extracted from image transforms and multi-order transforms can be found in \citep{Sha08,shamir2008evaluation,shamir2009knee,shamir2010impressionism,shamir2013automatic}.

The comprehensive nature of the numerical image content descriptors allows analyzing complex morphology such as radiology \citep{shamir2009knee,shamir2009early}, microscopy \citep{shamir2008iicbu,man14}, and visual art \citep{shamir2010impressionism,shamir2012pollock}. In particular, the Wndchrm feature set has been proved to be informative for analysis of galaxy morphology, and was found useful for tasks such as galaxy classification \citep{sha09, kum14} and automatic detection of peculiar galaxies \citep{shamir2012automatic,shamir2014automatic,shamir2014chloe}. A complete and detailed description of the set of numerical content descriptors and comprehensive performance analysis is available in \citep{Sha08,orlov2008wnd,shamir2008evaluation,shamir2010impressionism,shamir2009knee,shamir2009early}, and the source code is also publicly available through the Astrophysics Source Code Library \citep{shamir2013wnd}.

As mentioned in Section~\ref{introduction}, the purpose of the method is not to automatically classify galaxies by their morphology, but to quantitatively deduce a network of similarities between the different morphological types using merely the galaxy images, and without using metadata or existing knowledge that is not in the image content. The unsupervised analysis \citep{shamir2010impressionism,shamir2012computer,shamir2013automatic} works by first allocating 140 galaxy images from each galaxy type as annotated by EFIGI to the training set, and assigning each numerical image content descriptor with its Fisher discriminant score \citep{bishop2006pattern} computed using the training samples. After the content descriptors were ranked based on their Fisher discriminant scores, the 85\% of the least informative features, with the lowest Fisher scores, are rejected \citep{sha09,shamir2009knee,shamir2009evaluation}.

The similarity between each pair of galaxy images is then computed using the Weighted Nearest Distance (WND) algorithm \citep{Sha08,shamir2010impressionism}. The mean similarity between all test galaxies of type $t_1$ and all training galaxies of type $t_2$ determines the similarity between these two galaxy morphological types. The similarities between all pairs of galaxy types produce a similarity matrix, normalized such that the similarity between a certain type to itself is set to 1 \citep{Sha08,shamir2010impressionism,shamir2012computer,shamir2013automatic}. The similarity matrix is computed 20 times such that in each run different images are randomly allocated to training and test sets, and the final similarity matrix is generated by averaging the 20 similarity matrices.

The similarity matrix is visualized by PHYLIP \citep{felsenstein1993phylip,kuhner1994simulation}, which was originally developed for visualizing similarities between organisms by their genotypes, but in this experiment used to visualize the similarities between galaxy types. It is used with randomized input order of sequences where 97 is the seed, 10 jumbles, and the Equal-Daylight arc optimization. When pairs of nodes are added, new nodes are created to provide the optimal tree that represents the similarity matrix. PHYLIP first creates the tree in the form of a text file that follows the Newick format, and then visualizes it by using the DRAWTREE program. The edges between the nodes reflect the degree of similarities between them, such that a shorter path between two nodes reflects a higher similarity between the images of these two classes. DRAWTREE automatically sets the angles such that the tree is convenient and easy to read. In the phylogeny created by PHYLIP each pair of nodes has just one possible path between them, and the length of the path includes all segments on that path, including edges between nodes added by PHYLIP during the tree optimization process.


The method used to compute and visualize the similarities between the galaxy types is described in details in \citep{shamir2010impressionism,shamir2012computer,shamir2014classification}, and was used for unsupervised analysis of simulated images of galaxy mergers \citep{shamir2013automatic}. It also demonstrated its ability to profile continuous biomedical processes in which the clinical stages are reflected by image morphologies that change on a continuous scale \citep{sha2010prog}. Detailed instructions including specific command lines used to produce the results are described in~\ref{replication}.

\section{Results}
\label{results}

\begin{figure*}[ht]
\begin{center}
\includegraphics[scale=0.75]{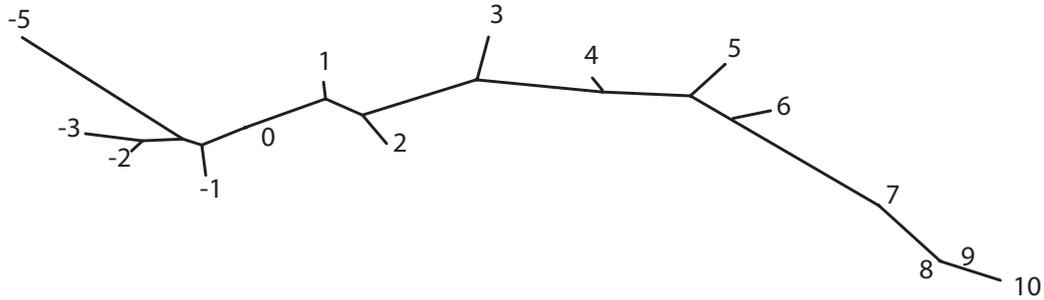}
\caption{The network of similarities between the galaxy morphological types as deduced automatically by the algorithm}
\label{color_sequence}
\end{center}
\end{figure*}

The application of the similarity estimation method described in Section~\ref{method} to the EFIGI color image data described in Section~\ref{data} produced the phylogeny displayed by Figure~\ref{color_sequence}.

As the figure shows, the network of similarities between the galaxy morphological types computed by the algorithm is in agreement with the ordered sequence proposed by \cite{de59}. The algorithm produced a graph starting with the ellipticals (-5), followed by the lenticualr galaxies (-3 through -1).  Then, continuing sequentially are the spiral galaxies from (1 through 9) followed by the irregulars (10) in an order with perfect agreement with \citep{de59}. As mentioned in Section~\ref{data}, the cD (-4), cE (-6), and dE types (11) were not included in the analysis due to the insufficient amount of sample images of these types in EFIGI. The probability that 15 elements are ordered in an ascending or descending order by mere chance is $\frac{2}{15!} = \sim 1.53\cdot10^{-12}$.

\begin{figure}[ht]
\begin{center}
\includegraphics[scale=0.45]{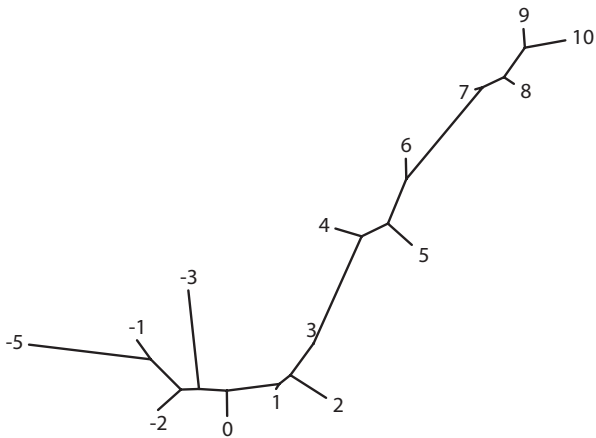}
\caption{The network of similarities between the galaxy types using normalized gray-scale images}
\label{bw_sequence}
\end{center}
\end{figure}

In another experiment we tested the method using the color images converted to gray-scale, and normalized for intensity such that all images had mean pixel value of 100, and standard deviation of 25 \citep{Sha08}. The normalization ensured that the order will be determined by the shape, with no impact of color or brightness. The resulting graph produced by the algorithm is displayed in Figure~\ref{bw_sequence}.

As the figure shows, the analysis of the normalized gray-scale images provided results similar to the graph produced using the color images, showing that the order was not necessarily driven by pixel intensity or by the color. The random chance probability that 12 elements out of 15 are ordered in ascending or descending order is  ${2 \cdot {15 \choose 12}} \frac{1}{12!} = \sim1.9\cdot10^{-6}$. 

It is also noticeable that the S0 galaxy types S0$^{-}$ (-3), S0$^{0}$ (-2), and S0$^{+}$ (-1) do not follow the numerical order proposed by \cite{de59}. That analysis of the computer is in agreement with the observation that S0$^{-}$, S0$^{0}$, and S0$^{+}$ galaxies do not feature an increase in the flattening of the galaxies \citep{san61}.

\begin{figure*}[ht]
\begin{center}
\includegraphics[scale=0.55]{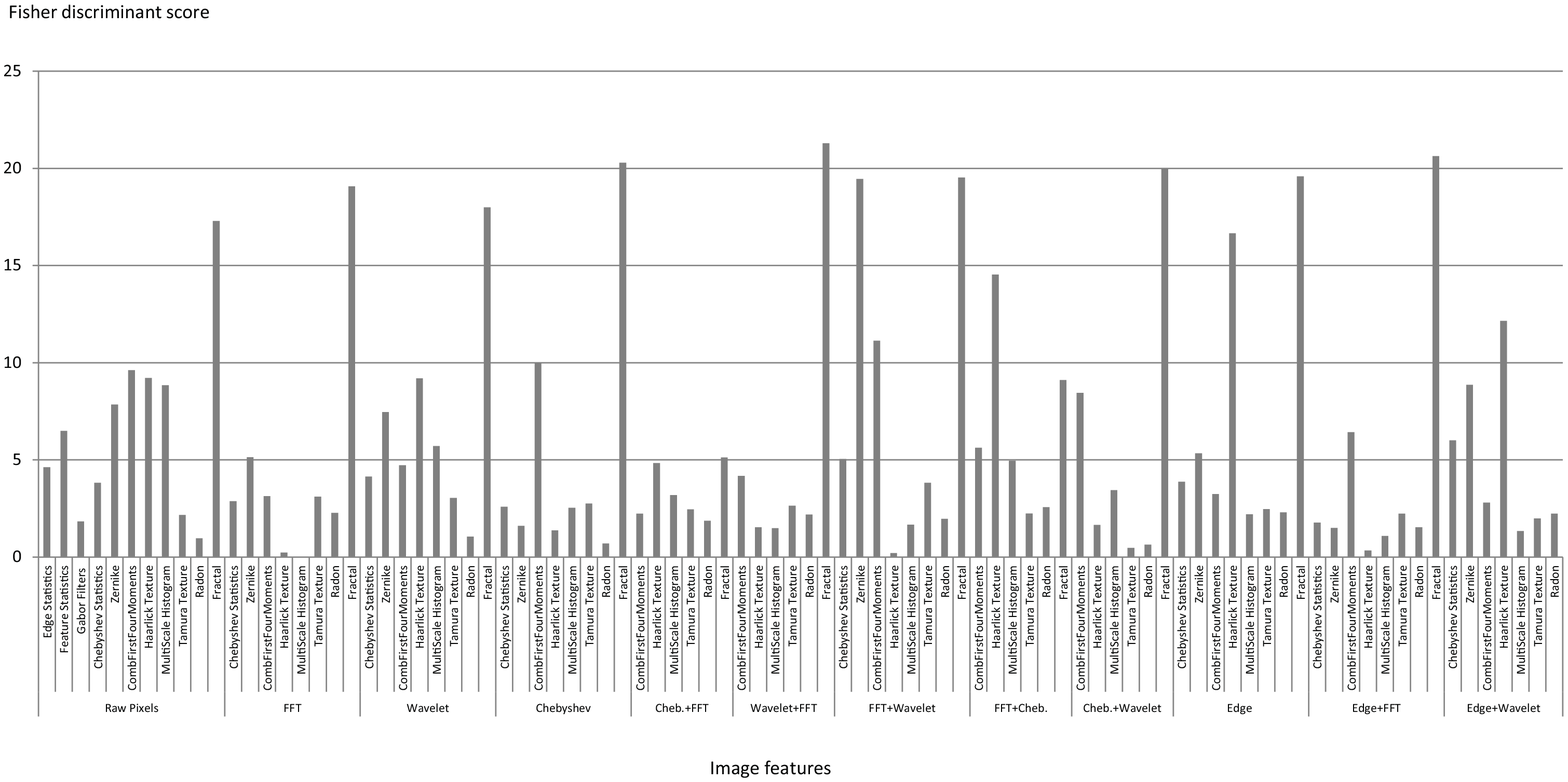}
\caption{Fisher discriminant scores of the different groups of numerical content descriptors, extracted from the different image transforms}
\label{features}
\end{center}
\end{figure*}

Figure~\ref{features} shows the Fisher discriminant scores of the groups of numerical image content descriptors, reflecting the measured informativeness of the descriptors and consequently their impact on the analysis. The descriptors are extracted from the image transforms and multi-order transforms.

As the figure shows, the identification of the Hubble stage depends on numerous image content descriptors working in concert. The fractal features were the most informative descriptors, indicating that the fractality of the galaxy is different across different galaxy morphological types. This agrees with the observation that fractality can be used as a galaxy classification signature \citep{lekshmi2003galaxy}, and can assist in differentiating between elliptical and spiral galaxies \citep{sha09}. For instance, an elliptical galaxy has low fractality in the absence of complex shape, but the fractality of a galaxy should become more dominant when the galaxy has more arms and split arms.

The graph shows that many other numerical image content descriptors such as Haralick textures \citep{haralick1973textural} have an impact on the analysis, and work in concert to quantify the similarities between the different galaxy morphological types. Texture features has been shown to be informative in separating between galaxies based on their morphological types \citep{au2006,sha09,banerji2010,pedersen2013}. For instance, texture homogeneity/entropy may change as the galaxy becomes more sparse, and the texture also correlates with star formation rate \citep{pedersen2013}.

On the other hand, several numerical content descriptors did not show substantial difference between galaxies of different Hubble stages. For instance, Radon features do not show a change between different galaxy types, as well as Tamura textures. The weak ability of Tamura textures to differentiate between galaxy types is that the directionality can be offset by galaxies or arms rotating to the opposite direction. That is different from other texture analysis algorithms such as Haralick, where the texture entropy and energy are independent of the direction.

\begin{figure*}[ht]
\begin{center}
\includegraphics[scale=0.50]{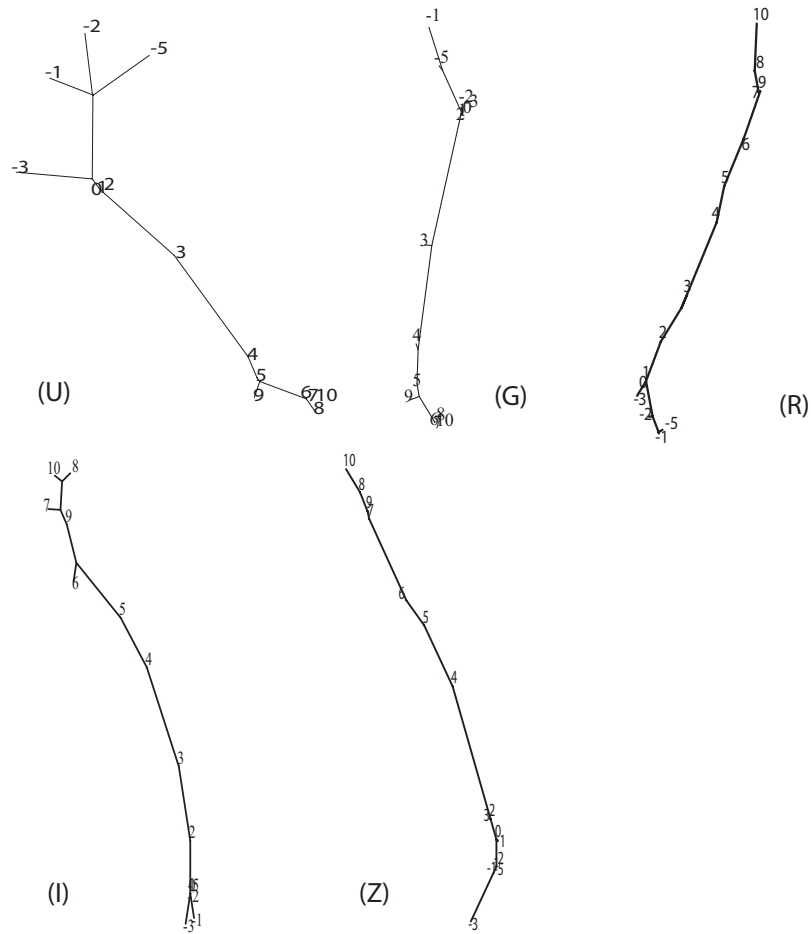}
\caption{The network of similarities between the galaxy types using the u, g, r, i, and z bands of EFIGI}
\label{u_g_r_i_z}
\end{center}
\end{figure*}

The experiment was also repeated with the EFIGI galaxy images of the u, g, i, r, and z bands. The resulting phylogenies are displayed by Figure~\ref{u_g_r_i_z}.

As the figure shows, the order of the galaxy types somewhat violates the \cite{de59} scheme. The shorter segments between some of the galaxy types show higher similarity deduced by the method, indicating that in some cases the algorithm could not identify the differences between these types. That shows that although the order of the galaxy types deduced by the algorithm is largely in agreement with the sequence described in \citep{de59}, processing just one band leads to loss of information, and consequently the order and automatic placement of the galaxy types is not as close to the order of \cite{de59} compared to the color images. Color has been identified to correlate with galaxy types \citep{str01}, and therefore color information can contribute to the ability of the algorithm to analyze the similarities between different types of galaxies. The probability to have the orders of the tree of the u, g, r, i, and z filters by chance is $6.84\cdot10^{-5}$, $6.84\cdot10^{-5}$, $1.9\cdot10^{-6}$, $1.6\cdot10^{-3}$, $6.84\cdot10^{-5}$, respectively.

Also, the analysis of the u band shows strong separation between late type galaxies and the other galaxy types, where Sa and Sab are positioned close to the lenticular galaxies. A similar observation can be made with the analysis of galaxies in the g band. The r, i, and z bands show a more even distribution of the types along the main axis, but it is also noticeable that the early types are clustered on one side of the axis, while irregular and Sd galaxies are grouped close to each other on the other side.

Figure~\ref{u_g_r_i_z} also shows that the early type galaxies could not be ordered correctly by the algorithm without using the color images, and that the individual bands or grayscale images did not have sufficient morphological features of these galaxy types that allow the automatic positioning in the same order proposed by \cite{de59}.

\section{Conclusion}
\label{conclusion}

Although galaxy classification cannot be considered a goal in itself \citep{de1994global}, it is a key to understanding the physical properties of the past, present, and future universe. Numerous galaxy morphological schemes have been proposed by manual observation and measurement of galaxy morphology and photometry. Here we proposed a computer-based approach to galaxy morphology by using an unsupervised machine learning system that can deduce the visual similarities between sets of images and reconstruct morphological sequences of galaxies. The analysis is performed such that the algorithm determines the network of similarities between the different morphological classes automatically, and without human guidance. 

The results show that when using the color EFIGI galaxy images the sequence deduced by the computer is in large agreement with the \cite{de59} scheme, even when using the color images as gray-scale images. When using each band separately the deduced order was in weaker agreement with \citep{de59}, showing that the composite color images contained more visual information that was used by the algorithm to deduce the order of the morphological types. The saturation and gamma correction applied to the EFIGI color images as described in Section~\ref{data} could also affect the way these images were analyzed. Basic statistical analysis shows very low probability of $\sim 1.53\cdot10^{-12}$ for having the galaxy types ordered in an ascending or descending order by mere chance.

The color images allowed the algorithm to deduce a sequence that is more consistent with the order of \cite{de59} compared to the sequences produced with each of the individual bands, indicating that the color images contained more information that was used by the algorithm to deduce the order of the morphological typess.

One difference between the \cite{de59} scheme and the network of morphological similarities produced by the algorithm  is the S0 galaxies, where the computer algorithm did not find the exact same order identified by \cite{de59}. The ability of the computer to deduce a network of similarities that is largely in agreement with manual analysis demonstrates the discovery power of the method, and its potential ability to analyze larger datasets containing a higher number of galaxy classes and identify and profile a possible morphological sequence. That allows quantitative morphological of entire galaxies, rather than the quantification of individual identifiable morphological features (e.g., the number of spiral arms).

While the experiments described in this paper are focused on galaxies in the Hubble sequence, with the increasing importance of digital sky surveys imaging billions of galaxies such as the Large Synoptic Survey Telescope (LSST), automated methods are also important to identify and analyze peculiar galaxies that cannot be associated with a defined morphological stage on the Hubble sequence. The scheme of numerical image content descriptors described in Section~\ref{method} has demonstrated its efficacy in detecting peculiar galaxy mergers among millions of galaxies in Sloan Digital Sky Survey, and performing quantitative assessment of these mergers \citep{shamir2014automatic}. Sky surveys such as LSST will be able to image a much larger number of galaxies, from which peculiar galaxies can be detected. Automatic detection methods such as \citep{shamir2012automatic,shamir2014automatic} can assist in the detection of peculiar galaxies that are not associated with stages on the Hubble sequence, and analysis methods such as the method described here can be used to identify links between a large number of galaxy types.

Source code of the analysis methods used in the experiment are publicly available, as well as the protocol as described in \ref{replication}.


\bibliographystyle{elsarticle-harv}
\bibliography{ms}

\newpage
\appendix
\section{Protocol}
\label{replication}

All software tools used to produce the results are open source, making it easier to replicate the results or analyze other datasets \citep{shamir2013practices}. Source code for the computer analysis method \citep{Sha08} is available at the Astrophysics Source Code Library \citep{shamir2013wnd} or at http://vfacstaff.ltu.edu/lshamir/downloads/ImageClassifier, as well as its dependency libraries  \citep{shamir2013wnd}. It also requires the installation of the open source PHYLIP package, available at http://evolution.genetics.washington.edu/phylip.html. The experiments also require computational resources that can process the EFIGI catalog. The experiment in this paper was done with a 16-core Intel Core-i7 machine and 32GB of RAM, and took about three days to complete.

To replicate the results, the following steps are required: \newline \newline
1. Download the EFIGI catalog from http://www.astromatic.net/projects/efigi \newline \newline
2. Convert the color FITS images (or PNG images) to TIF format by using ImageMagick. A batch conversion can be done by the following command line: find /path/to/efigi -name ``*.FITS'' -exec convert \{\} \{\}.tif $\backslash$; \newline \newline
3. Separate the images into folders such that the name of each folder is the T number, and its content is the galaxy images of that T number as annotated by EFIGI. \newline \newline
4. Compute the image features by running the command line:  ./wndchrm train -mlc /path/to/efigi\_root\_folder /path/to/efigi\_root\_folder/efigi.fit \newline That step might take several days to complete with a single core, but the response time can be shorter by running several instances of the process. The process should not be stopped to avoid the creation of empty .sig files. In case the process stopped before completion, the following command line should be used before running it again:  find /path/to/efigi -name ``*.sig'' -exec rm \{\} $\backslash$; \newline \newline
5. The phylogeny can be created by running the following command line: ./wndchrm test -f0.15 -i\#140 -j12 -n20 -w -p/path/to/phylip /path/to/efigi\_root\_folder/efigi.fit /path/to/efigi\_root\_folder/efigi.html \newline When done, a .ps file should be created in the folder ``/path/to/efigi\_root\_folder''. \newline \newline
6. To process the grayscale images step 4 should be replaced with the command line:  ./wndchrm train -ml -S100:25 /path/to/efigi\_root\_folder /path/to/efigi\_root\_folder/efigi.fit \newline \newline

The experiments were performed in Linux (Fedora) environment. For further information or assistance please contact the authors.

%
%
\end{document}